\documentclass[aps,prd,preprint,superscriptaddress,showpacs,floatfix,nobibnotes,nofootinbib]{revtex4}
\usepackage{mathrsfs}

\usepackage{bm}
\usepackage{graphicx}
\usepackage{longtable}
\usepackage{amsmath}
\usepackage{amssymb}

\begin{document}

\title{Unstable molecular state with unstable constituent}

\author{Yuqin Yin}
\affiliation{Department of Fundamental Courses, Shandong University of Science and Technology, Taian 271019, China}

\author{Zonghua Shi}
 \affiliation{Department of Fundamental Courses, Shandong University of Science and Technology, Taian 271019, China}

\author{Hao Liu}
 \affiliation{Department of Fundamental Courses, Shandong University of Science and Technology, Taian 271019, China}

\author{Xiurong Guo}
 \affiliation{Department of Fundamental Courses, Shandong University of Science and Technology, Taian 271019, China}

\author{Xiaozhao Chen}\email{chen_xzhao@sina.com}
\email[corresponding author]{} \affiliation{Department of Fundamental Courses, Shandong University of Science and Technology, Taian 271019, China}

\author{Xiaofu L\"{u}}
\affiliation{Department of Physics, Sichuan University, Chengdu 610064, China}
\affiliation{Institute of Theoretical Physics, The Chinese Academy of Sciences, Beijing 100080, China}
\affiliation{CCAST (World Laboratory), P.O. Box 8730, Beijing  100080, China}

\date{\today}

\begin{abstract}
Based on the developed Bethe-Salpeter theory for dealing with unstable state, we investigate unstable meson-meson molecular state in which at least one of the constituents is an unstable meson and provide a reasonable and feasible scheme to deal with this interesting problem in the framework of relativistic quantum field theory. The developed Bethe-Salpeter theory is applied to investigate the unstable meson composed of a quark and an antiquark, and we obtain the Green's function for unstable meson in the framework of relativistic quantum field theory, which is used to deal with unstable constituent of molecular state. We can obtain the physical mass and width of unstable molecular state composed of two heavy mesons, which contain the contribution from at least one unstable constituent of molecular state.
\end{abstract}

\pacs{12.40.Yx, 14.40.Rt, 12.39.Ki}


\maketitle

\newpage

\parindent=20pt

\section{introduction}
Hadronic molecule structure has been proposed to interpret the internal structure of exotic meson resonance for many years \cite{ms:Swanson,ms:Torn}. However, the previous works seldom took into account
unstable constituent of molecular state in the framework of relativistic quantum field theory \cite{ms:Guo,ms:Dong,ms:Liu,fourbody1,fourbody2,fourbody3,fourbody4,mypaper8,mypaper9}. In Refs. \cite{mypaper8,mypaper9}, we developed Bethe-Salpeter (BS) theory for dealing with unstable two-body system and provided a reasonable and feasible scheme to describe unstable system in the framework of relativistic quantum field theory. In this work, we will apply the developed Bethe-Salpeter theory (DBST) to investigate unstable constituent of molecular state.

In this work, we consider that exotic meson resonance is an unstable molecular state composed of two heavy mesons and that at least one of the constituents is a narrow resonance. DBST can be applied to investigate the unstable meson composed of a quark and an antiquark. According to DBST, we obtain the Green's function for unstable composite particle in the framework of relativistic quantum field theory, which is used to deal with unstable constituent of molecular state.

In DBST, we suppose that at some given time the unstable two-body system has been prepared to decay and then study the time evolution of this system as determined by the total Hamiltonian. Since the prepared state has not decayed, we can consider that the heavy mesons in molecular state have not decayed and these constituents of molecular state can be regarded as stable particles composed of quark and antiquark. Then the prepared state can be considered as a bound state composed of two stable composite particles, and this prepared state can be described by the ground-state BS wave function for bound state at the given time. From the effective theory at low energy QCD, we construct the interaction kernel between two quarks in two quark-antiquark bound states derived from one light-meson exchange and one heavy-meson exchange in the framework of relativistic quantum field theory. Solving BS equation with this interaction kernel, we can obtain the mass and BS wave function for bound state composed of two stable quark-antiquark bound states. After providing the description for the prepared state, we can study the time evolution of the prepared state and obtain the pole corresponding to resonance through the scattering matrix element.

The crucial point of DBST is that the scattering matrix element is calculated in the framework of relativistic quantum field theory. According to DBST, the scattering matrix element should be calculated with respect to arbitrary value of the final state energy. The total energy of the final state extends over the real interval while the initial state energy is specified \cite{mypaper8}. In this work, we consider unstable constituent of molecular state, and the scattering matrix element should contain the contribution from unstable constituent of molecular state. Because the constituent meson of molecular state is narrow resonance, the propagator for the constituent meson should be replaced by the corresponding Green's function in the framework of relativistic quantum field theory. Using the Green's function for unstable composite particle, we can obtain the scattering matrix element which contains the contribution from unstable constituent of molecular state. Finally, we obtain the correction for energy level of molecular state due to decay channels and decay widths of exotic meson resonance, which contain the contribution from at least one unstable constituent of molecular state.

The structure of this article is as follows. Section \ref{sec:DBST} gives the developed Bethe-Salpeter theory. In Sec. \ref{sec:BS} we give BS equation for bound state composed of two quark-antiquark bound states, which is considered as the prepared state. The mass and BS wave function for bound state composed of two quark-antiquark bound states are obtained. In Sec. \ref{sec:T-ME} we introduce the Green's function for unstable composite particle in the framework of relativistic quantum field theory, which is used to deal with unstable constituent of molecular state. The matrix elements for at least one unstable constituent of molecular state with respect to arbitrary value of the final state energy are obtained. In Sec. \ref{sec:mw} we obtain the physical mass and width for unstable molecular state. We make some concluding remarks in Sec. \ref{sec:concl}.

\section{the developed Bethe-Salpeter theory}\label{sec:DBST}
To deal with unstable composite particle in the framework of relativistic quantum field theory, we considered the time evolution of unstable two-body system as determined by the total Hamiltonian and provided the developed Bethe-Salpeter theory in Refs. \cite{mypaper8,mypaper9}. According to the developed Bethe-Salpeter theory for dealing with unstable two-body system \cite{mypaper8}, this unstable system has been prepared to decay at given time, and the prepared state can be regarded as a bound state with ground-state energy. Solving BS equation for arbitrary two-body bound state, one can obtain the mass $M_0$ and BS wave function $\chi_P(x_1,x_2)$ for this bound state with momentum $P=(\mathbf{P},i\sqrt{\mathbf{P}^2+M_0^2})$, where $x_1=(\mathbf{x}_1,it_1)$ and $x_2=(\mathbf{x}_2,it_2)$. Setting $t_1=0$ and $t_2=0$ in the ground-state BS wave function, we obtain a description for the prepared state (ps)
\begin{equation}
\begin{split}\label{BSWFT0}
\mathscr{X}^{\text{ps}}_a=\chi_{P}(\mathbf{x}_1,t_1=0,\mathbf{x}_2,t_2=0)=\frac{1}{(2\pi)^{3/2}}\frac{1}{\sqrt{2E(P)}}e^{i\mathbf{P}\cdot(\eta_1\mathbf{x}_1+\eta_2\mathbf{x}_2)}\chi_P(\mathbf{x}_1-\mathbf{x}_2),
\end{split}
\end{equation}
where $E(p)=\sqrt{\mathbf{p}^2+m^2}$ and $\eta_1+\eta_2=1$. We emphasize that the prepared state is not the physical state and the prepared state mass $M_0$ is not the physical mass of unstable composite particle.

Now it is necessary to consider the total Hamiltonian
\begin{equation}
\begin{split}
H=K_I+V_I,
\end{split}
\end{equation}
where $K_I$ represents the interaction responsible for the formation of stationary bound state and $V_I$ stands for the interaction responsible for the decay of unstable composite particle. Then the time evolution of this system determined by the total Hamiltonian $H$ has the explicit form
\begin{equation}
\begin{split}\label{timeevo}
\mathscr{X}(t)=e^{-iHt}\mathscr{X}^{\text{ps}}_{a}=\frac{1}{2\pi i}\int_{C_2}d\epsilon e^{-i\epsilon t}\frac{1}{\epsilon-H}\mathscr{X}^{\text{ps}}_{a},
\end{split}
\end{equation}
where $(\epsilon-H)^{-1}$ is the operator for the Green's function and the contour $C_2$ runs from $ic_r+\infty$ to $ic_r-\infty$ in energy-plane. The positive constant $c_r$ is sufficiently large that no singularity of $(\epsilon-H)^{-1}$ lies above $C_2$. The time-dependent wave function $\mathscr{X}(t)$ provides a complete description of the system for $t>0$. Since $H\neq K_I$, this system should not remain in the prepared state $\mathscr{X}^{\text{ps}}_a$. Then at arbitrary time $t$ the probability amplitude of finding the system in the state $\mathscr{X}^{\text{ps}}_a$ is
\begin{equation}
\begin{split}
\mathscr{A}_a=(\mathscr{X}^{\text{ps}}_{a},\mathscr{X}(t))=\frac{1}{2\pi i}\int_{C_2}d\epsilon\frac{ie^{-i\epsilon t}}{\epsilon-M_0-(2\pi)^3T_{aa}(\epsilon)}.
\end{split}
\end{equation}
In field theory the operator $T(\epsilon)$ is just the scattering matrix with energy $\epsilon$, and $T_{aa}(\epsilon)$ is the $T$-matrix element between two bound states, which is defined as
\begin{equation}
\begin{split}
\langle a~\text{out}|a~\text{in}\rangle=\langle a~\text{in}|a~\text{in}\rangle-i(2\pi)^4\delta^{(4)}(P-P)T_{aa}(\epsilon).
\end{split}
\end{equation}

Because of the analyticity of $T_{aa}(\epsilon)$, we define
\begin{equation}
\begin{split}\label{Taepsilon}
T_{aa}(\epsilon)=\mathbb{D}(\epsilon)-i\mathbb{I}(\epsilon),
\end{split}
\end{equation}
where $\epsilon$ approaches the real axis from above, $\mathbb{D}$ and $\mathbb{I}$ are the real and imaginary parts, respectively. Using the unitarity of $T_{aa}(\epsilon)$, we obtain \cite{GreenFun}
\begin{equation}
\begin{split}\label{Taepsilon1}
2\mathbb{I}(\epsilon)=\sum_{c'}\sum_b(2\pi)^4\delta^{(3)}(\mathbf{P}_b-\mathbf{P})\delta(E_b-\epsilon)|T_{(c';b)a}(\epsilon)|^2,
\end{split}
\end{equation}
where $P_b=(\mathbf{P}_b,iE_b)$ is the total energy-momentum vector of all particles in the final state and the $T$-matrix element $T_{(c';b)a}(\epsilon)$ is defined as $\langle (c';b)~\text{out}|a~\text{in}\rangle=-i(2\pi)^4\delta^{(3)}(\mathbf{P}_b-\mathbf{P})\delta(E_b-\epsilon)T_{(c';b)a}(\epsilon)$ in channel $c'$. The delta-function in Eq. (\ref{Taepsilon1}) means that the energy $\epsilon$ in scattering matrix is equal to the total energy $E_b$ of the final state, $\sum_b$ represents summing over momenta and spins of all particles in the final state, and $\sum_{c'}$ represents summing over all possible channels. For $E_b=\epsilon$, we also denote the total energy of the final state by $\epsilon$ and $\mathbb{I}(\epsilon)$ becomes a function of the final state energy. Using dispersion relation for the function $T_{aa}(\epsilon)$, we obtain
\begin{equation}
\begin{split}\label{disrel}
\mathbb{D}(\epsilon)=-\frac{\mathcal{P}}{\pi}\int_{\epsilon_M}^\infty \frac{\mathbb{I}(\epsilon')}{\epsilon'-\epsilon}d\epsilon',
\end{split}
\end{equation}
where the symbol $\mathcal{P}$ means that this integral is a principal value integral and the variable of integration is the total energy $\epsilon'$ of the final state. To calculate the real part, we need calculate the function $\mathbb{I}(\epsilon')$ of value of the final state energy $\epsilon'$, which is an arbitrary real number over the real interval $\epsilon_M<\epsilon'<\infty$. As usual the momentum of initial bound state $a$ is set as $P=(0,0,0,iM_0)$ in the rest frame and $\epsilon_M$ denotes the sum of all particle masses in the final state. We suppose that the final state $b$ may contain $n$ composite particles and $n'$ elementary particles in decay channel $c'$. From Eq. (\ref{Taepsilon1}), we have
\begin{equation}
\begin{split}\label{Iepsilon'}
\mathbb{I}(\epsilon')=&\frac{1}{2}\sum_{c'}\int d^3Q'_1\cdots d^3Q'_{n'}d^3Q_1\cdots d^3Q_n(2\pi)^4\delta^{(4)}(Q'_1+\cdots+Q_n-P^{\epsilon'})\sum_{\text{spins}}|T_{(c';b)a}(\epsilon')|^2,
\end{split}
\end{equation}
where $\sum_{c'}$ represents summing over all open and closed channels, $Q'_1\cdots Q'_{n'}$ and $Q_1\cdots Q_n$ are the momenta of final elementary and composite particles, respectively; $P^{\epsilon'}=(0,0,0,i\epsilon')$, $T_{(c';b)a}(\epsilon')$ is the $T$-matrix element with respect to $\epsilon'$, and $\sum_{\text{spins}}$ represents summing over spins of all particles in the final state. In Eq. (\ref{Iepsilon'}) the energy in scattering matrix is equal to the total energy $\epsilon'$ of the final state $b$, which is an arbitrary real number over the real interval $\epsilon_M<\epsilon'<\infty$. Because the total energy $\epsilon'$ of the final state extends from $\epsilon_M$ to $+\infty$, we may obtain several closed channels derived from the interaction Lagrangian. The mass $M_0$ and BS amplitude of initial bound state $a$ have been specified and the value of the initial state energy in the rest frame is a specified value $M_0$. From Eq. (\ref{Iepsilon'}), we have $\mathbb{I}(\epsilon')>0$ for $\epsilon'>\epsilon_M$ and $\mathbb{I}(\epsilon')=0$ for $\epsilon'\leqslant\epsilon_M$, which is the reason that the integration in dispersion relation (\ref{disrel}) ranges from $\epsilon_M$ to $+\infty$.

In experiments, many unstable particles are narrow states and their decay widths are very small compared with their energy levels, i.e., $(2\pi)^3\mathbb{I}(M_0)\ll M_0$. This situation is ordinarily interpreted as implying that both $(2\pi)^3|\mathbb{D}(\epsilon)|$ and $(2\pi)^3\mathbb{I}(\epsilon)$ are also very small quantities, as compared to $M_0$. Therefore, we can expect that $[\epsilon-M_0-(2\pi)^3T_{aa}(\epsilon)]^{-1}$ has a pole on the second Riemann sheet
\begin{equation}
\begin{split}\label{pole}
\epsilon_{\text{pole}}\cong M_0+(2\pi)^3[\mathbb{D}(M_0)-i\mathbb{I}(M_0)]=M-i\frac{\Gamma(M_0)}{2},
\end{split}
\end{equation}
where $\Delta M=(2\pi)^3\mathbb{D}(M_0)$ is the correction for energy level of resonance and $M=M_0+(2\pi)^3\mathbb{D}(M_0)$ is the physical mass for resonance. This pole at $\epsilon_{\text{pole}}$ describes the resonance. The mass $M_0$ of two-body bound state is obtained by solving homogeneous BS equation, which should not be the mass of physical resonance. $\Gamma(M_0)$ with mass $M_0$ also should not be the width $\Gamma$ of physical resonance, which should depend on its physical mass $M$. Then we obtain the Green's function for unstable composite particle in the rest frame
\begin{equation}
\begin{split}\label{Greenfun}
G(\epsilon)=\frac{i}{\epsilon-M+i\Gamma/2},
\end{split}
\end{equation}
where the contribution from negative energy is not considered.

\section{the prepared state}\label{sec:BS}
At low energy QCD, an important feature is the spontaneous breaking of chiral symmetry, and the quark-gluon coupling constant becomes very large. In the actual calculation perturbation theory is thoroughly inadequate and it is impossible to exactly solve BS equation from the QCD Lagrangian. Fortunately, based on the spontaneous breaking of chiral symmetry, the effective theory at low energy QCD has been constructed \cite{QTFII}. According to the effective theory at low energy QCD, non-vanishing vacuum condensate causes the spontaneous breaking of chiral symmetry, which leads to the appearance of Goldstone bosons. The current quark becomes constituent quark. At low energy QCD, the effective interaction Lagrangian can be regarded as Lagrangian for the interaction of light mesons with constituent quarks. In this paper, we investigate the light meson interaction with the light quarks in heavy mesons and the interaction Lagrangian for the coupling of constituent quark fields to light meson fields is \cite{mypaper6}
\begin{equation}\label{Lag}
\begin{split}
&\mathscr{L}^{\text{eff}}_I=ig_0\left(\begin{array}{ccc} \bar{u}&\bar{d}&\bar{s}
\end{array}\right)\gamma_5\left(\begin{array}{ccc}
\pi^0+\frac{1}{\sqrt{3}}\eta &\sqrt{2}\pi^+&\sqrt{2}K^+\\\sqrt{2}\pi^-&-\pi^0+\frac{1}{\sqrt{3}}\eta &\sqrt{2}K^0\\\sqrt{2}K^-&\sqrt{2}\bar{K}^0&-\frac{2}{\sqrt{3}}\eta
\end{array}\right)\left(\begin{array}{c} u\\d\\s
\end{array}\right)\\
&+ig'_0\left(\begin{array}{ccc} \bar{u}&\bar{d}&\bar{s}
\end{array}\right)\gamma_\mu\left(\begin{array}{ccc}
\rho^0+\omega&\sqrt{2}\rho^+&\sqrt{2}K^{*+}\\\sqrt{2}\rho^-&-\rho^0+\omega&\sqrt{2}K^{*0}\\\sqrt{2}K^{*-}&\sqrt{2}\bar{K}^{*0}&\sqrt{2}\phi
\end{array}\right)_\mu\left(\begin{array}{c} u\\d\\s
\end{array}\right)+g_\sigma\left(\begin{array}{cc} \bar{u}&\bar{d}
\end{array}\right)\left(\begin{array}{c} u\\d
\end{array}\right)\sigma.
\end{split}
\end{equation}
From this effective interaction Lagrangian at low energy QCD, we have to consider that heavy meson is composed of a quark and an antiquark and investigate the interaction of light meson with constituent quarks in heavy meson.

Here, we imagine that an exotic meson resonance is an unstable molecular state composed of two heavy vector mesons ($VM$ and $\overline{VM'}$) and that this exotic resonance decays into a heavy axial-vector meson ($HM$) and a light pseudoscalar meson ($LM$). Moreover, the constituents ($VM$ and $\overline{VM'}$) of molecular state may be unstable particles. As mentioned in Sec. \ref{sec:DBST}, there are two steps to deal with this unstable system. As the first step, we investigate stable bound state, which is considered as the prepared state. As the second step, we study the time evolution of unstable system determined by the total Hamiltonian and obtain the correction for energy level of resonance due to decay channels. In this section, our attention is only focused on the prepared state. Since the prepared state has not decayed, we can consider that the heavy mesons ($VM$ and $\overline{VM'}$) in molecular state have not decayed and these constituents of molecular state can be regarded as stable particles composed of quark and antiquark. In this section, heavy vector mesons are considered as stable quark-antiquark bound states and the prepared state is a bound state ($MS$) composed of two stable vector particles.

If a bound state with spin $j$ and parity $\eta_{P}$ is created by two Heisenberg vector fields with masses $M_1$ and $M_2$, respectively, its BS wave function is defined as
\begin{equation}\label{BSwfdx}
\chi_{P(\lambda\tau)}^j(x_1',x_2')=\langle0|TA_\lambda(x_1')A^\dagger_\tau(x_2')|P,j\rangle=\frac{1}{(2\pi)^{3/2}}\frac{1}{\sqrt{2E(P)}}e^{iP\cdot X}\chi_{P(\lambda\tau)}^j(X'),
\end{equation}
where $P$ is the momentum of the bound state, $X=\eta_1x_1'+\eta_2x_2'$, $X'=x_1'-x_2'$ and $\eta_{1,2}=M_{1,2}/(M_1+M_2)$. Making the Fourier transformation, we obtain BS wave function in the momentum representation
\begin{equation}\label{BSwfdp}
\begin{split}
\chi^j_P(p_1',p_2')_{\lambda\tau}=\frac{1}{(2\pi)^{3/2}}\frac{1}{\sqrt{2E(P)}}(2\pi)^4\delta^{(4)}(P-p_1'+p_2')\chi^j_{\lambda\tau}(P,p),
\end{split}
\end{equation}
where $p$ is the relative momentum of two vector fields and we have $P=p_1'-p_2'$, $p=\eta_2p_1'+\eta_1p_2'$, $p_1'$ and $p_2'$ are the momenta carried by two vector fields, respectively. In Ref. \cite{mypaper9}, we have given the general form of BS wave functions for the bound states created by two massive vector fields with arbitrary spin and definite parity, for $\eta_{P}=(-1)^j$,
\begin{equation}\label{jp}
\begin{split}
\chi_{\lambda\tau}^{j}(P,p)=&\frac{1}{\mathcal{N}^j}\eta_{\mu_1\cdots\mu_j}[p_{\mu_1}\cdots p_{\mu_j}(\mathcal{T}^1_{\lambda\tau}\Phi_1+\mathcal{T}^2_{\lambda\tau}\Phi_2)+\mathcal{T}^3_{\mu_1\cdots\mu_j\lambda\tau}\Phi_3+\mathcal{T}^4_{\mu_1\cdots\mu_j\lambda\tau}\Phi_4\\
&+\mathcal{T}^5_{\mu_1\cdots\mu_j\lambda\tau}\Phi_5+\mathcal{T}^6_{\mu_1\cdots\mu_j\lambda\tau}\Phi_6],
\end{split}
\end{equation}
for $\eta_{P}=(-1)^{j+1}$,
\begin{equation}\label{jm}
\begin{split}
\chi_{\lambda\tau}^{j}(P,p)=&\frac{1}{\mathcal{N}^j}\eta_{\mu_1\cdots\mu_j}(p_{\mu_1}\cdots p_{\mu_j}\epsilon_{\lambda\tau\xi\zeta}p_\xi P_\zeta\Phi'_1+\mathcal{T}^7_{\mu_1\cdots\mu_j\lambda\tau}\Phi'_2+\mathcal{T}^8_{\mu_1\cdots\mu_j\lambda\tau}\Phi'_3+\mathcal{T}^{9}_{\mu_1\cdots\mu_j\lambda\tau}\Phi'_4\\
&+\mathcal{T}^{10}_{\mu_1\cdots\mu_j\lambda\tau}\Phi'_5+\mathcal{T}^{11}_{\mu_1\cdots\mu_j\lambda\tau}\Phi'_6+\mathcal{T}^{12}_{\mu_1\cdots\mu_j\lambda\tau}\Phi'_7),
\end{split}
\end{equation}
where $\mathcal{N}^j$ is normalization, $\eta_{\mu_1\cdots\mu_j}$ is the polarization tensor describing the spin of the bound state, subscripts $\lambda$ and $\tau$ are derived from these two vector fields, the independent tensor structures $\mathcal{T}^i_{\lambda\tau}$ are given in Appendix \ref{app1}, $\Phi_i(P\cdot p,p^2)$ and $\Phi'_i(P\cdot p,p^2)$ are independent scalar functions.

BS wave function describing bound state composed of two stable heavy vector particles should satisfy BS equation
\begin{equation}\label{BSE}
\chi^{j}_{\lambda\tau}(P,p)=-\int \frac{d^4p'}{(2\pi)^4}\Delta_{F\lambda\theta}(p_1')\mathcal{V}_{\theta\theta',\kappa'\kappa}(p,p';P)\chi^{j}_{\theta'\kappa'}(P,p')\Delta_{F\kappa\tau}(p_2'),
\end{equation}
where $\mathcal{V}_{\theta\theta',\kappa'\kappa}$ is the interaction kernel, $P=(0,0,0,iM_0)$, $p_1'=p+P/2$, $p_2'=p-P/2$, $\Delta_{F\lambda\theta}(p_1')$ and $\Delta_{F\kappa\tau}(p_2')$ are the propagators for the spin 1 fields, $\Delta_{F\lambda\theta}(p_1')=(\delta_{\lambda\theta}+\frac{p'_{1\lambda} p'_{1\theta}}{M_1^2})\frac{-i}{p_1'^2+M_1^2-i\varepsilon}$, $\Delta_{F\kappa\tau}(p_2')=(\delta_{\kappa\tau}+\frac{p'_{2\kappa} p'_{2\tau}}{M_2^2})\frac{-i}{p_2'^2+M_2^2-i\varepsilon}$. We emphasize that the kernel $\mathcal{V}$ is defined in two-body channel so $\mathcal{V}$ is not complete interaction. The kernel in homogeneous BS equation (\ref{BSE}) plays a central role for making two-body system to be a stable bound state, and the solution of homogeneous BS equation (\ref{BSE}) should only describe bound state. In our approach, BS equation for meson-meson bound state is treated in the ladder approximation. This approximation consists in replacing the interaction kernel by its lowest order value corresponding to the simple one-meson exchange.

In this section, the heavy mesons in a molecular state are considered as bound states composed of a heavy quark and a light quark. From the interaction Lagrangian for the coupling of light quark fields to light meson fields expressed as Eq. (\ref{Lag}), we can obtain the interaction kernel between two light quarks in two heavy mesons from one light meson exchange. Moreover, we should consider the interaction kernel between two heavy quarks in two heavy mesons from one heavy meson exchange. In our theoretical frame, the interaction kernel between two heavy mesons is derived from one meson exchange between two quarks in these two heavy mesons. To construct the interaction kernel between two stable heavy vector particles, we consider one light meson exchange and one heavy meson exchange. Using the approach introduced in Refs. \cite{mypaper3,mypaper4,mypaper5,mypaper9}, we can obtain the interaction kernel from one light meson exchange and one heavy meson exchange, and the details for constructing the interaction kernel are shown in Ref. \cite{mypaper9}. Then we can solve BS equation (\ref{BSE}) and the mass $M_0$ and BS wave function $\chi^{j}_{\lambda\tau}(P,p)$ of bound state composed of two stable heavy vector particles can be obtained. The reduced normalization condition for $\chi_{\lambda\tau}^{j}(P,p)$ is
\begin{equation}
\begin{split}
&\frac{-i}{(2\pi)^4}\int d^4p\bar\chi^j_{\lambda'\tau'}(P,p)\frac{\partial}{\partial P_0}[\Delta_{F\lambda'\lambda}(p+P/2)^{-1}\Delta_{F\tau\tau'}(p-P/2)^{-1}]\chi^j_{\lambda\tau}(P,p)=(2P_0)^2,
\end{split}
\end{equation}
where $\Delta_{F\beta\alpha'}(p)^{-1}$ is the inverse propagator for the vector field with mass $m$, $\Delta_{F\beta\alpha'}(p)^{-1}=i(\delta_{\beta\alpha'}-\frac{p_{\beta}p_{\alpha'}}{p^2+m^2})(p^2+m^2)$ \cite{mypaper6}. BS amplitude of bound state composed of two stable heavy vector particles is
\begin{equation}
\begin{split}
\Gamma^j_{\theta\kappa}(P,p)=\Delta_{F\theta\theta'}(p_1')^{-1}\chi^j_{\theta'\kappa'}(P,p)\Delta_{F\kappa'\kappa}(p_2')^{-1}.
\end{split}
\end{equation}

In this section, the heavy meson is considered as a bound state consisting of a quark and an antiquark and the meson-meson bound state is actually a four-quark state. The generalized BS wave function for four-quark state describing the bound state composed of two stable heavy vector particles has the form \cite{fourbody1,mypaper6,mypaper7}
\begin{equation}\label{fourquarkBSWF1}
\begin{split}
\chi^{j}(P,p,k,k')=\chi_{\lambda}(p_1',k)\Delta_{F\lambda\theta}(p_1')\Gamma^j_{\theta\kappa}(P,p)\Delta_{F\kappa\tau}(p_2')\chi_\tau(p_2',k'),
\end{split}
\end{equation}
where $\chi_{\lambda}(p_1',k)$ and $\chi_\tau(p_2',k')$ are BS wave functions of two stable heavy vector mesons, respectively; $k$ and $k'$ are the relative momenta between quark and antiquark in these two mesons. BS wave functions of heavy vector mesons have been given in Refs. \cite{BSE:Roberts1,BSE:Roberts3,BSE:Roberts4,BSE:Roberts5,BSE:Krassnigg}. Then setting $x_1'=(\mathbf{x}_1',0)$ and $x_2'=(\mathbf{x}_2',0)$, we obtain BS wave function of the prepared state. It is necessary to emphasize that the prepared state is not the physical state and the prepared state mass $M_0$ is not the physical mass of resonance.

\section{$T$-matrix element $T_{(c';b)a}(\epsilon')$}\label{sec:T-ME}
In order to obtain the real part $\mathbb{D}(\epsilon)$ due to all decay channels, we have to calculate $T_{(c';b)a}(\epsilon')$ with respect to value of the final state energy $\epsilon'$, which is an arbitrary real number over the real interval $\epsilon_M<\epsilon'<\infty$. Here, we imagine that an exotic meson resonance is an unstable molecular state composed of two heavy vector mesons ($VM$ and $\overline{VM'}$) and that this exotic resonance decays into a heavy meson ($HM$) and a light meson ($LM$). This decay mode is denoted as $c'_1$. When calculating the $T$-matrix element $T_{(c'_1;b)a}(\epsilon')$, we emphatically consider that at least one of the constituents ($VM$ and $\overline{VM'}$) of molecular state is an unstable particle in this section.

\subsection{Green's function for unstable constituent}\label{sec:Green'fun}
DBST can be applied to investigate the unstable meson composed of a quark and an antiquark. It is straightforward from Eq. (\ref{Greenfun}) to identify the Green's function with momentum $p$ as
\begin{equation}
\begin{split}\label{Greenfun1}
G(p)=\frac{i}{p_0-\sqrt{\mathbf{p}^2+(M_{\text{um}}-i\Gamma_{\text{um}}/2)^2}},
\end{split}
\end{equation}
where the momentum $4$-vector is $p=(\mathbf{p},ip_{0})$, $M_{\text{um}}$ and $\Gamma_{\text{um}}$ are the physical mass and width of unstable meson, respectively. Considering the contribution from negative energy, we obtain the Green's function for unstable scalar meson in the framework of relativistic quantum field theory
\begin{equation}
\begin{split}\label{Greenfun2}
G(p)=&\frac{i}{[p_0-\sqrt{\mathbf{p}^2+(M_{\text{um}}-i\Gamma_{\text{um}}/2)^2}][p_0+\sqrt{\mathbf{p}^2+(M_{\text{um}}-i\Gamma_{\text{um}}/2)^2}]}\\
=&\frac{-i}{p^2+(M_{\text{um}}-i\Gamma_{\text{um}}/2)^2}.
\end{split}
\end{equation}
Similarly, the Green's function for unstable vector meson has the form
\begin{equation}
\begin{split}\label{Greenfun3}
G_{\mu\nu}(p)=&\bigg(\delta_{\mu\nu}+\frac{p_\mu p_\nu}{M^2_{\text{um}}}\bigg)\frac{-i}{p^2+(M_{\text{um}}-i\Gamma_{\text{um}}/2)^2}.
\end{split}
\end{equation}
In this work, we use the corresponding Green's function to deal with unstable constituent of molecular state.

\subsection{Channel $c'_1$}\label{sec:hcpie'}
In our approach, an unstable molecular state is composed of two heavy vector mesons $VM$ and $\overline{VM'}$, and these two heavy vector mesons may be unstable composite particles consisting of a quark and an antiquark. Owing to the effective interaction Lagrangian at low energy QCD (\ref{Lag}), we consider that in the final state the light pseudoscalar meson $LM$ is an elementary particle and the heavy axial-vector meson $HM$ is a bound state of a heavy quark and a heavy antiquark.

Firstly, we present our previous method to deal with the heavy mesons in a molecular state. In our previous works \cite{mypaper8,mypaper9}, the heavy mesons in a molecular state were considered as bound states composed of a heavy quark and a light quark. Using Mandelstam's approach, we have calculated the matrix elements between bound states. Mandelstam's approach is a technique based on BS wave function for evaluating the general matrix element between bound states. Applying Mandelstam's approach, one can express the general matrix element between bound states in terms of BS wave functions and a two-particle irreducible Green's function \cite{ParticlesFields}. In DBST \cite{mypaper8}, Mandelstam's approach has been generalized to evaluate the bound state matrix element with respect to arbitrary value of the final state energy $\epsilon'$.

\begin{figure}[!htb] \centering
\includegraphics[trim = 0mm 30mm 0mm 30mm,scale=1,width=10cm]{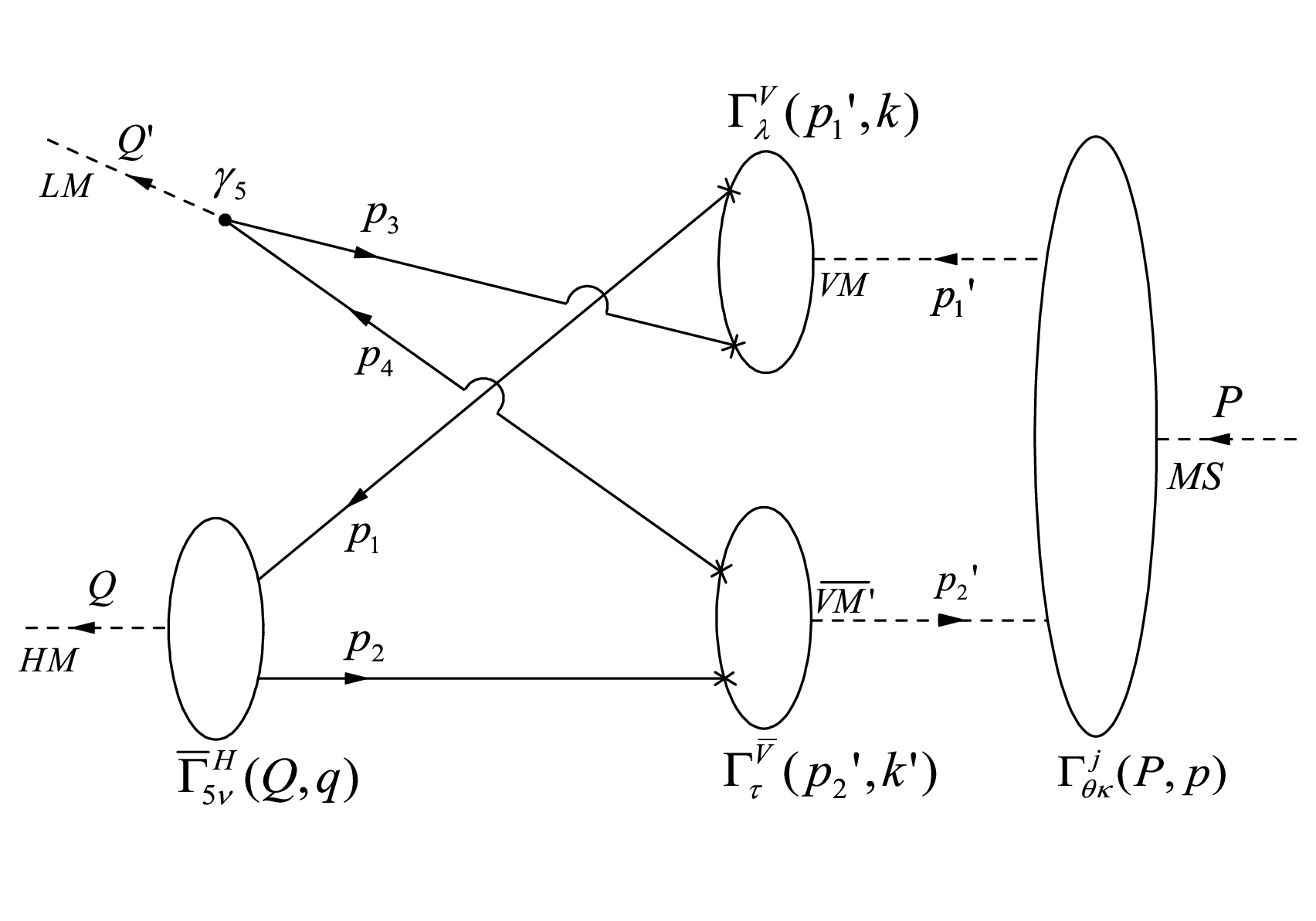}
\caption{\label{Fig1} Matrix element with respect to $\epsilon'$ for channel $c'_1$. The solid lines denote quark propagators, and the unfilled ellipses represent Bethe-Salpeter amplitudes. The momenta in the final state satisfy $Q+Q'=P^{\epsilon'}$ and the momentum of the initial state is $P$. The final state energy extends from $\epsilon_{c'_1,M}$ to $+\infty$ while the initial state energy is specified, and the crosses mean that the momenta of quark propagators depend on the final state energy $\epsilon'$.}
\end{figure}

Applying the generalized Mandelstam's approach and retaining the lowest order term of the two-particle irreducible Green's function, we can obtain the $T$-matrix element $T_{(c'_1;b)a}(\epsilon')$ with respect to arbitrary value of the final state energy $\epsilon'$ for channel $c'_1$ in the momentum representation
\begin{equation}
\begin{split}\label{Tmatrele1}
T_{(c'_1;b)a}(\epsilon')=&\frac{-i}{(2\pi)^{9/2}}\frac{g_0}{\sqrt{2E_{L}(Q')}}\frac{\varepsilon_\nu^{\varrho}(Q)}{\sqrt{2E_{H}(Q)}}\frac{1}{\sqrt{2E(P)}}\int \frac{d^4kd^4p}{(2\pi)^8}\\
&\times\text{Tr}[\bar\chi^{H}_{5\nu}(Q,q)\Gamma^{V}_\lambda(p_1',k)S_F(p_3)\gamma_5 S_F(p_4)\Gamma^{{\bar V}}_\tau(p_2',k')\Delta_{F\lambda\theta}(p_1')\Gamma^j_{\theta\kappa}(P,p)\Delta_{F\kappa\tau}(p_2')],
\end{split}
\end{equation}
where the total energy $\epsilon'$ of the final state extends from $\epsilon_{c'_1,M}$ to $+\infty$, i.e., $\epsilon_{c'_1,M}<\epsilon'<\infty$; $Q$ and $Q'$ represent the momenta of final particles ($HM$ and $LM$), $Q^2=-M_{H}^2$ and $Q'^2=-M_{L}^2$; $E(p)=\sqrt{\mathbf{p}^2+m^2}$; $p_1, p_3, p_4, p_2$ are the momenta of four quarks; $p_1'$ and $p_2'$ are the momenta of two heavy vector mesons; $q$, $k$ and $k'$ are the relative momenta between quark and antiquark in heavy mesons, respectively; $\varepsilon_\nu^{\varrho=1,2,3}(Q)$ is the polarization vector of $HM$ in the final state, $S_F(p)$ is the quark propagator, $\Gamma_\lambda^{V}(K,k)$ represents BS amplitude of heavy vector meson $VM$, $\chi^H_{5\nu}(Q,q)$ represents BS wave function for heavy axial-vector meson $HM$ in the final state, and $\Gamma^j_{\theta\kappa}(P,p)$ represents BS amplitude for the initial bound state ($MS$) composed of two stable heavy vector particles. For heavy mesons, the authors of Refs. \cite{BSE:Roberts1,BSE:Roberts3,BSE:Roberts4,BSE:Roberts5,BSE:Krassnigg} have obtained their BS amplitudes. It is necessary to emphasize that the energy in the two-particle irreducible Green's function is equal to the final state energy $\epsilon'$ while the mass $M_0$ and BS amplitude of initial bound state are specified. In our previous method \cite{mypaper9}, the heavy mesons in a molecular state were considered as quark-antiquark bound states, so the generalized BS amplitude of initial meson-meson bound state should be $\Gamma^{V}_\lambda(p_1',k)\Delta_{F\lambda\theta}(p_1')\Gamma^j_{\theta\kappa}(P,p)\Delta_{F\kappa\tau}(p_2')\Gamma^{{\bar V}}_\tau(p_2',k')$, which has been specified. We have introduced \emph{extended Feynman diagram} in Ref. \cite{mypaper8} to represent arbitrary value of the final state energy, shown as Fig. \ref{Fig1}. In Fig. \ref{Fig1}, the quark momenta in left-hand side of crosses depend on the final state energy and the momenta in right-hand side depend on the initial state energy, i.e., $p_1-p_2-p_3+p_4=Q+Q'=P^{\epsilon'}$ and $p_1'-p_2'=P$. In the rest frame, we have $P=(0,0,0,iM_0)$, $P^{\epsilon'}=(0,0,0,i\epsilon')$, $\epsilon_{c'_1,M}<\epsilon'<\infty$ and $\epsilon_{c'_1,M}=M_{H}+M_L$. When $\epsilon'=M_0$, the crosses in Fig. \ref{Fig1} disappear and then the extended Feynman diagram becomes the traditional Feynman diagram. These momenta should become
\begin{equation}\label{momenta}
\begin{split}
&p_1=(Q+Q')/2+p+k,~~p_2=(Q+Q')/2-Q+p+k,~~p_3=k,~~p_4=Q'+k,\\
&q=Q'/2+p+k,~k'=Q'(M_0)+k,~~p_1'=p+P/2,~~p_2'=p-P/2,~~Q+Q'=P^{\epsilon'},
\end{split}
\end{equation}
where $Q'(M_0)=(-\mathbf{Q}(M_0),i\sqrt{\mathbf{Q}^2(M_0)+M_L^2})$, $Q=(\mathbf{Q}(\epsilon'),i\sqrt{\mathbf{Q}^2(\epsilon')+M_{H}^2})$, $Q'=(-\mathbf{Q}(\epsilon'),i\sqrt{\mathbf{Q}^2(\epsilon')+M_L^2})$, $\mathbf{Q}^2(M_0)=[M_0^2-(M_{H}+M_L)^2][M_0^2-(M_{H}-M_L)^2]/(4M_0^2)$ and $\mathbf{Q}^2(\epsilon')=[\epsilon'^2-(M_{H}+M_L)^2][\epsilon'^2-(M_{H}-M_L)^2]/(4\epsilon'^2)$.

\begin{figure}[tbp]
\centering 
\includegraphics[width=.49\textwidth,clip]{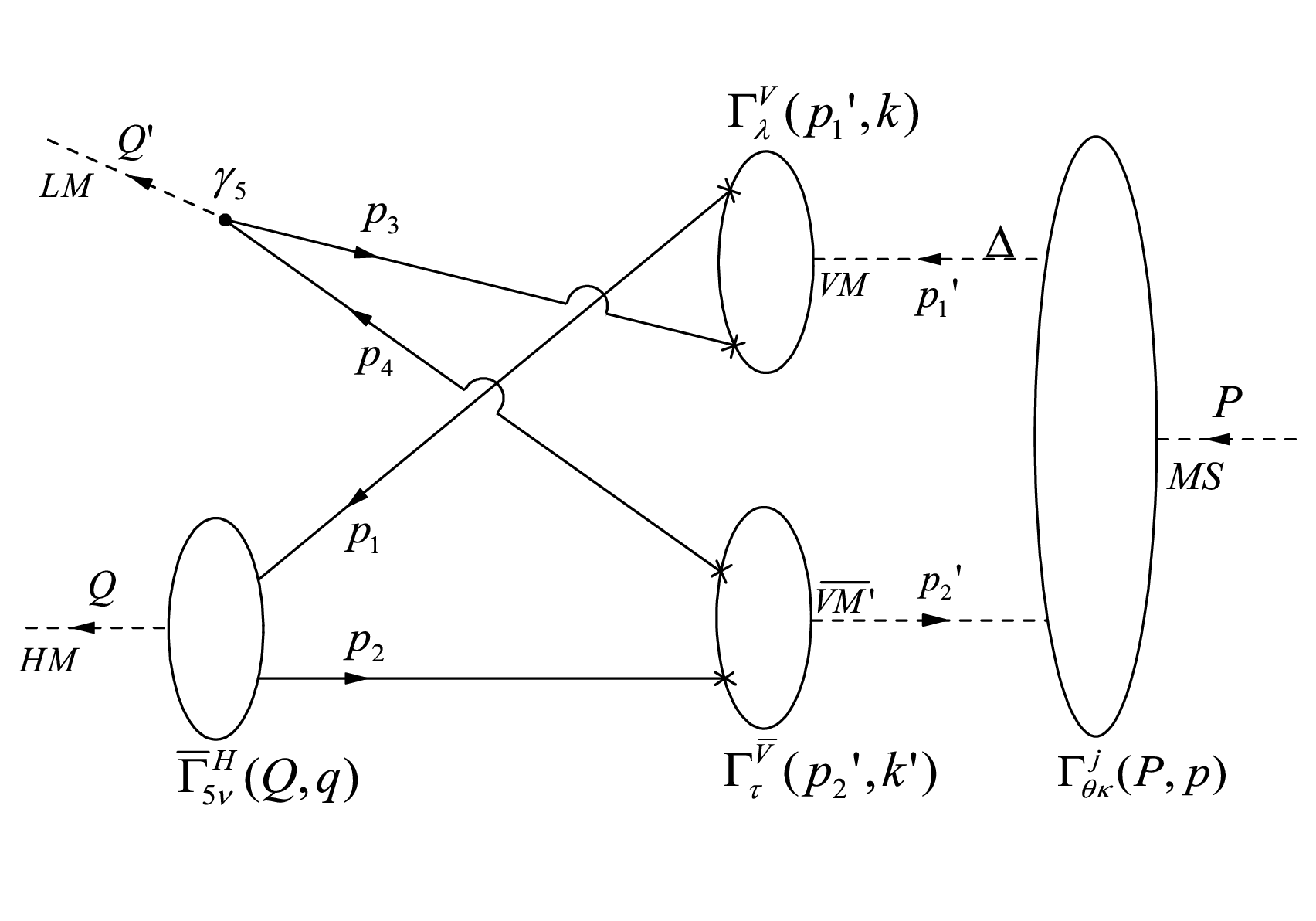}
\hfill
\includegraphics[width=.49\textwidth,origin=c]{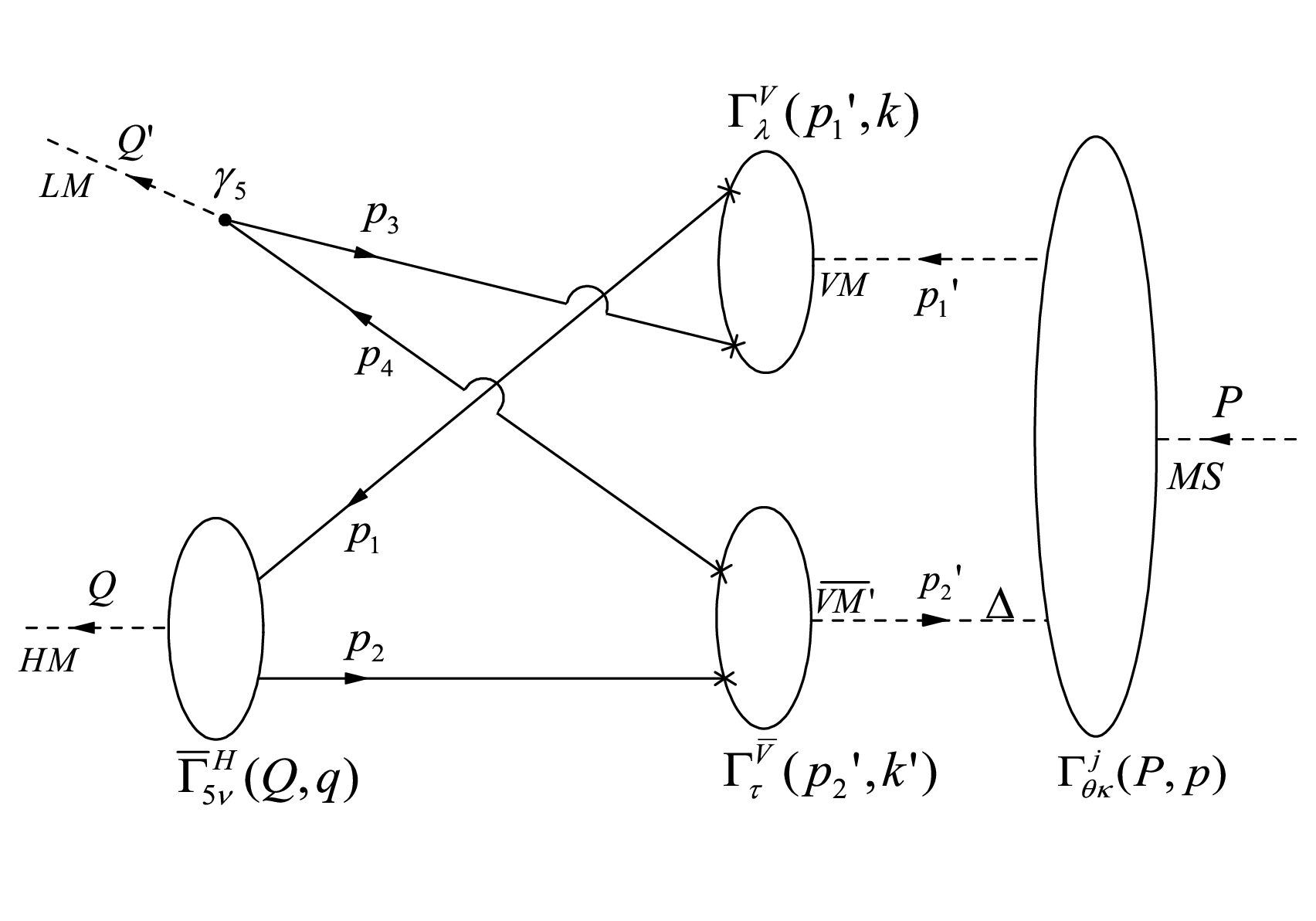}
\caption{\label{Fig2} Matrix element for one unstable constituent ($VM$ or $\overline{VM'}$) of molecular state. The dashed line with $\Delta$ means the Green's function for unstable vector meson.}
\end{figure}

\begin{figure}[!htb] \centering
\includegraphics[trim = 0mm 30mm 0mm 30mm,scale=1,width=10cm]{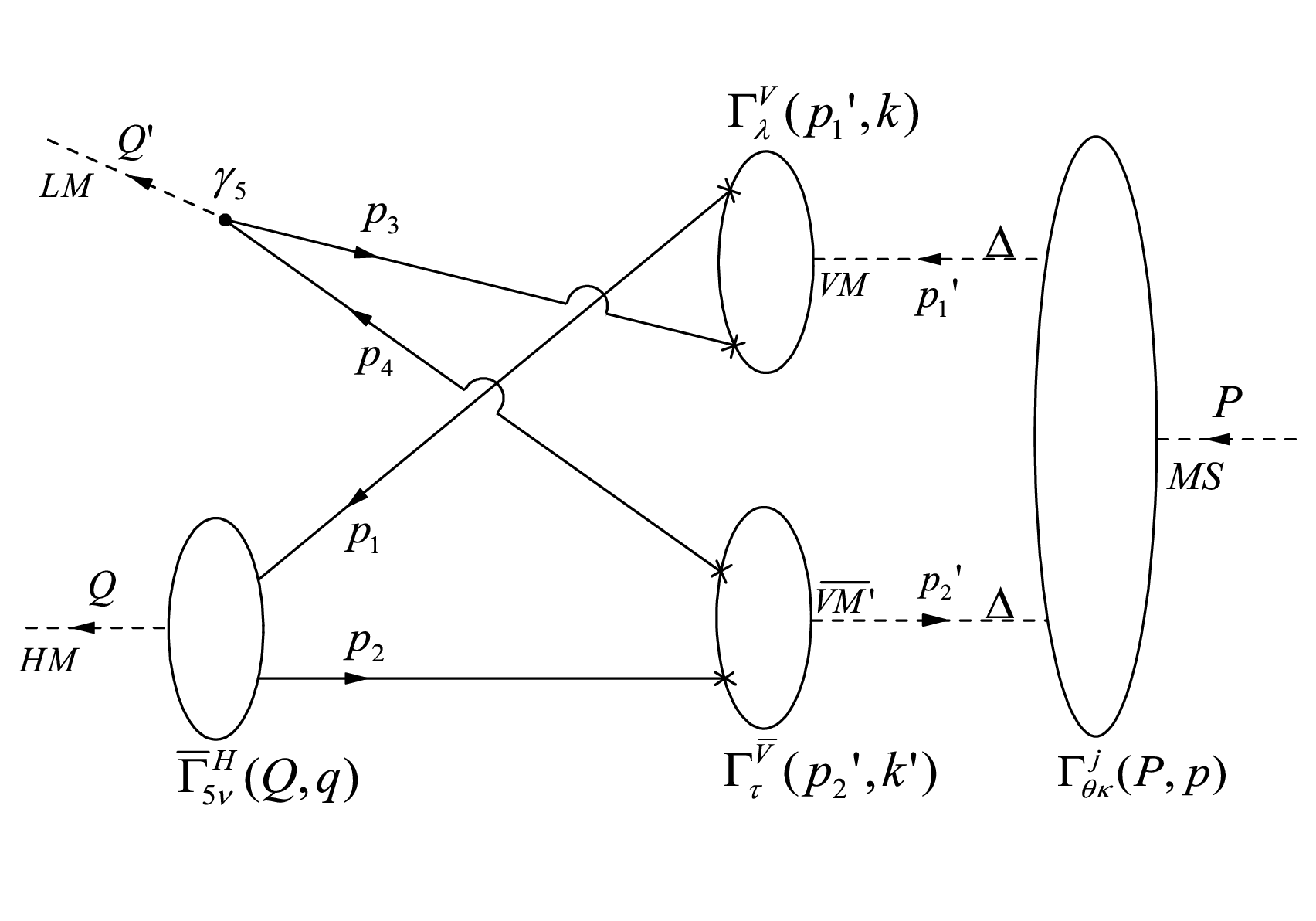}
\caption{\label{Fig3} Matrix element for both unstable constituents ($VM$ and $\overline{VM'}$) of molecular state. The dashed line with $\Delta$ means the Green's function for unstable vector meson.}
\end{figure}

In the previous work, we did not consider that the constituents ($VM$ and $\overline{VM'}$) of molecular state may be unstable composite particles. In this work, we provide a reasonable and feasible scheme to deal with this interesting problem in the framework of relativistic quantum field theory as follows. Because the constituent meson of molecular state is unstable, the propagator for the constituent meson should be replaced by the corresponding Green's function. If one of the constituents ($VM$ and $\overline{VM'}$) of molecular state is unstable composite particle, Eq. (\ref{Tmatrele1}) becomes
\begin{equation}
\begin{split}\label{Tmatrele2a}
T_{(c'_1;b)a}(\epsilon')=&\frac{-i}{(2\pi)^{9/2}}\frac{g_0}{\sqrt{2E_{L}(Q')}}\frac{\varepsilon_\nu^{\varrho}(Q)}{\sqrt{2E_{H}(Q)}}\frac{1}{\sqrt{2E(P)}}\int \frac{d^4kd^4p}{(2\pi)^8}\\
&\times\text{Tr}[\bar\chi^{H}_{5\nu}(Q,q)\Gamma^{V}_\lambda(p_1',k)S_F(p_3)\gamma_5 S_F(p_4)\Gamma^{{\bar V}}_\tau(p_2',k')G_{\lambda\theta}(p_1')\Gamma^j_{\theta\kappa}(P,p)\Delta_{F\kappa\tau}(p_2')],
\end{split}
\end{equation}
or
\begin{equation}
\begin{split}\label{Tmatrele2b}
T_{(c'_1;b)a}(\epsilon')=&\frac{-i}{(2\pi)^{9/2}}\frac{g_0}{\sqrt{2E_{L}(Q')}}\frac{\varepsilon_\nu^{\varrho}(Q)}{\sqrt{2E_{H}(Q)}}\frac{1}{\sqrt{2E(P)}}\int \frac{d^4kd^4p}{(2\pi)^8}\\
&\times\text{Tr}[\bar\chi^{H}_{5\nu}(Q,q)\Gamma^{V}_\lambda(p_1',k)S_F(p_3)\gamma_5 S_F(p_4)\Gamma^{{\bar V}}_\tau(p_2',k')\Delta_{F\lambda\theta}(p_1')\Gamma^j_{\theta\kappa}(P,p)G_{\kappa\tau}(p_2')],
\end{split}
\end{equation}
where $G_{\lambda\theta}(p_1')$ and $G_{\kappa\tau}(p_2')$ represent the Green's functions for unstable vector mesons. The Green's function for unstable vector meson is expressed as Eq. (\ref{Greenfun3}). In this paper, the extended Feynman diagram contains the contribution from unstable constituent of molecular state, as shown in Fig. \ref{Fig2}. The dashed line with $\Delta$ in these extended Feynman diagrams means the Green's function for unstable vector meson. If both constituents of molecular state are unstable composite particles, Eq. (\ref{Tmatrele1}) becomes
\begin{equation}
\begin{split}\label{Tmatrele3}
T_{(c'_1;b)a}(\epsilon')=&\frac{-i}{(2\pi)^{9/2}}\frac{g_0}{\sqrt{2E_{L}(Q')}}\frac{\varepsilon_\nu^{\varrho}(Q)}{\sqrt{2E_{H}(Q)}}\frac{1}{\sqrt{2E(P)}}\int \frac{d^4kd^4p}{(2\pi)^8}\\
&\times\text{Tr}[\bar\chi^{H}_{5\nu}(Q,q)\Gamma^{V}_\lambda(p_1',k)S_F(p_3)\gamma_5 S_F(p_4)\Gamma^{{\bar V}}_\tau(p_2',k')G_{\lambda\theta}(p_1')\Gamma^j_{\theta\kappa}(P,p)G_{\kappa\tau}(p_2')].
\end{split}
\end{equation}
This is shown in Fig. \ref{Fig3}. From Eq. (\ref{Iepsilon'}), we obtain the function $\mathbb{I}_1(\epsilon')$ for channel $c'_1$
\begin{equation}\label{Iepsilon'1}
\begin{split}
\mathbb{I}_1(\epsilon')&=\frac{1}{2}\int d^3Qd^3Q'(2\pi)^4\delta^{(4)}(Q+Q'-P^{\epsilon'})\sum_{\varrho=1}^3|T_{(c'_1;b)a}(\epsilon')|^2.
\end{split}
\end{equation}
Obviously, the function $\mathbb{I}_1(\epsilon')$ also contains the contribution from unstable constituent of molecular state.

\section{physical mass and width of resonance}\label{sec:mw}
For unstable molecular state, the dispersion relation (\ref{disrel}) becomes
\begin{equation}
\begin{split}
\mathbb{D}(M_0)=&-\frac{\mathcal{P}}{\pi}\int_{\epsilon_{c'_1,M}}^\infty \frac{\mathbb{I}_1(\epsilon')}{\epsilon'-M_0}d\epsilon',
\end{split}
\end{equation}
where $\epsilon_{c'_1,M}=M_{H}+M_L$. From Eq. (\ref{pole}), we obtain that the physical mass of unstable molecular state composed of two heavy vector mesons is $M=M_0+(2\pi)^3\mathbb{D}(M_0)$, which contains the contribution from at least one unstable constituent of molecular state. Replacing $M_0$ by $M$ in Eqs. (\ref{Tmatrele2a}) and (\ref{Tmatrele2b}) and setting $\epsilon'=M$, we calculate the matrix element $T_{(c'_1;b)a}(M)$ for one unstable constituent of molecular state. Replacing $M_0$ by $M$ in Eq. (\ref{Tmatrele3}) and setting $\epsilon'=M$, we calculate the matrix element $T_{(c'_1;b)a}(M)$ for both unstable constituents of molecular state. Then we can obtain that the corresponding width of unstable molecular state is $\Gamma_1=2(2\pi)^3\mathbb{I}_1(M)$.

\section{conclusion}\label{sec:concl}
In the framework of relativistic quantum field theory, we investigate unstable meson-meson molecular state in which at least one of the constituents is an unstable meson and provide a reasonable and feasible scheme to deal with this interesting problem. The constituent meson of molecular state is considered as narrow resonance, and we apply the developed Bethe-Salpeter theory to investigate the unstable meson composed of a quark and an antiquark. According to DBST, we obtain the Green's function for unstable composite particle in the framework of relativistic quantum field theory, which is used to calculate the scattering matrix element with respect to arbitrary value of the final state energy. Then we obtain the physical mass and decay width of unstable molecular state composed of two heavy mesons, which contain the contribution from at least one unstable constituent of molecular state. Obviously, this work provides a further improvement for the molecular hypothesis.

\begin{acknowledgements}
This work was supported by the National Natural Science Foundation of China under Grants No. 11705104, No. 11801323 and No. 52174145; Shandong Provincial Natural Science Foundation, China under Grants No. ZR2023MA083, No. ZR2016AQ19 and No. ZR2016AM31; and SDUST Research Fund under Grant No. 2018TDJH101.
\end{acknowledgements}

\appendix

\section{Tensor structures in the general form of BS wave functions}\label{app1}
The tensor structures in Eqs. (\ref{jp}) and (\ref{jm}) are given below \cite{mypaper9}
\begin{equation*}
\mathcal{T}_{\lambda\tau}^1=(p^2+\eta_1P\cdot p-\eta_2P\cdot p-\eta_1\eta_2P^2)g_{\lambda\tau}-(p_{\lambda}p_{\tau}+\eta_1P_{\tau}p_{\lambda}-\eta_2P_{\lambda}p_{\tau}-\eta_1\eta_2P_{\lambda}P_{\tau}),
\end{equation*}
\begin{equation*}
\begin{split}
\mathcal{T}_{\lambda\tau}^2=&(p^2+2\eta_1P\cdot p+\eta_1^2P^2)(p^2-2\eta_2P\cdot p+\eta_2^2P^2)g_{\lambda\tau}\\
&+(p^2+\eta_1P\cdot p-\eta_2P\cdot p-\eta_1\eta_2P^2)(p_{\lambda}p_{\tau}+\eta_1P_{\lambda}p_{\tau}-\eta_2P_{\tau}p_{\lambda}-\eta_1\eta_2P_{\lambda}P_{\tau})\\
&-(p^2-2\eta_2P\cdot p+\eta_2^2P^2)(p_{\lambda}p_{\tau}+\eta_1P_{\lambda}p_{\tau}+\eta_1P_{\tau}p_{\lambda}+\eta_1 ^2P_{\lambda}P_{\tau})\\
&-(p^2+2\eta_1P\cdot p+\eta_1^2P^2)(p_{\lambda}p_{\tau}-\eta_2P_{\lambda}p_{\tau}-\eta_2P_{\tau}p_{\lambda}+\eta_2 ^2P_{\lambda}P_{\tau}),
\end{split}
\end{equation*}
\begin{equation*}
\begin{split}
\mathcal{T}_{\mu_1\cdots\mu_j\lambda\tau}^3=&\frac{1}{j!}p_{\{\mu_2}\cdots p_{\mu_j}g_{\mu_1\}\lambda}(p^2+2\eta_1P\cdot p+\eta_1^2P^2)[(p^2-2\eta_2P\cdot p+\eta_2^2P^2)(p+\eta_1P)_{\tau}\\
&-(p^2+\eta_1P\cdot p-\eta_2P\cdot p-\eta_1\eta_2P^2)(p-\eta_2P)_{\tau}]\\
&-p_{\mu_1}\cdots p_{\mu_j}[(p^2-2\eta_2P\cdot p+\eta_2^2P^2)(p_{\lambda}p_{\tau}+\eta_1P_{\lambda}p_{\tau}+\eta_1P_{\tau}p_{\lambda}+\eta_1 ^2P_{\lambda}P_{\tau})\\
&-(p^2+\eta_1P\cdot p-\eta_2P\cdot p-\eta_1\eta_2P^2)(p_{\lambda}p_{\tau}+\eta_1P_{\lambda}p_{\tau}-\eta_2P_{\tau}p_{\lambda}-\eta_1\eta_2P_{\lambda}P_{\tau})],
\end{split}
\end{equation*}
\begin{equation*}
\begin{split}
\mathcal{T}_{\mu_1\cdots\mu_j\lambda\tau}^4=&\frac{1}{j!}p_{\{\mu_2}\cdots p_{\mu_j}g_{\mu_1\}\tau}(p^2-2\eta_2P\cdot p+\eta_2^2P^2)[(p^2+\eta_1P\cdot p\\
&-\eta_2P\cdot p-\eta_1\eta_2P^2)(p+\eta_1P)_{\lambda}-(p^2+2\eta_1P\cdot p+\eta_1^2P^2)(p-\eta_2P)_{\lambda}]\\
&+p_{\mu_1}\cdots p_{\mu_j}[(p^2+2\eta_1P\cdot p+\eta_1^2P^2)(p_{\lambda}p_{\tau}-\eta_2P_{\lambda}p_{\tau}-\eta_2P_{\tau}p_{\lambda}+\eta_2 ^2P_{\lambda}P_{\tau})\\
&-(p^2+\eta_1P\cdot p-\eta_2P\cdot p-\eta_1\eta_2P^2)(p_{\lambda}p_{\tau}+\eta_1P_{\lambda}p_{\tau}-\eta_2P_{\tau}p_{\lambda}-\eta_1\eta_2P_{\lambda}P_{\tau})],
\end{split}
\end{equation*}
\begin{equation*}
\begin{split}
\mathcal{T}^5_{\mu_1\cdots\mu_j\lambda\tau}=&\frac{1}{j!}(p^2+2\eta_1P\cdot p+\eta_1^2P^2)(p^2-2\eta_2P\cdot p+\eta_2^2P^2)p_{\{\mu_3}\cdots p_{\mu_j}g_{\mu_1\lambda}g_{\mu_2\}\tau}\\
&-\frac{1}{j!}p_{\{\mu_2}\cdots p_{\mu_j}g_{\mu_1\}\tau}(p^2-2\eta_2P\cdot p+\eta_2^2P^2)(p+\eta_1P)_{\lambda}\\
&-\frac{1}{j!}p_{\{\mu_2}\cdots p_{\mu_j}g_{\mu_1\}\lambda}(p^2+2\eta_1P\cdot p+\eta_1^2P^2)(p-\eta_2P)_{\tau}\\
&+p_{\mu_1}\cdots p_{\mu_j}(p_{\lambda}p_{\tau}+\eta_1P_{\lambda}p_{\tau}-\eta_2P_{\tau}p_{\lambda}-\eta_1\eta_2P_{\lambda}P_{\tau}),
\end{split}
\end{equation*}
\begin{equation*}
\begin{split}
\mathcal{T}^6_{\mu_1\cdots\mu_j\lambda\tau}=&p_{\{\mu_3}\cdots p_{\mu_j}\epsilon_{\mu_{1}\lambda\xi\zeta}p_\xi P_\zeta\epsilon_{\mu_{2}\}\tau\xi'\zeta'}p_{\xi'}P_{\zeta'},
\end{split}
\end{equation*}
\begin{equation*}
\begin{split}
\mathcal{T}_{\mu_1\cdots\mu_j\lambda\tau}^7=&-(2p^2+\eta_1P\cdot p-\eta_2P\cdot p)p_{\{\mu_2}\cdots p_{\mu_j}\epsilon_{\mu_1\}\lambda\tau\xi}p_\xi\\
&+(2\eta_1\eta_2P\cdot p+\eta_2p^2-\eta_1p^2)p_{\{\mu_2}\cdots p_{\mu_j}\epsilon_{\mu_1\}\lambda\tau\xi}P_\xi\\
&+p_{\{\mu_2}\cdots p_{\mu_j}\epsilon_{\mu_1\}\lambda\xi\zeta}p_\xi P_\zeta p_\tau+p_{\{\mu_2}\cdots p_{\mu_j}\epsilon_{\mu_1\}\tau\xi\zeta}p_\xi P_\zeta p_\lambda,
\end{split}
\end{equation*}
\begin{equation*}\
\begin{split}
\mathcal{T}_{\mu_1\cdots\mu_j\lambda\tau}^8=&-(P\cdot p)p_{\{\mu_2}\cdots p_{\mu_j}\epsilon_{\mu_1\}\lambda\tau\xi}p_\xi+p^2p_{\{\mu_2}\cdots p_{\mu_j}\epsilon_{\mu_1\}\lambda\tau\xi}P_\xi\\
&-p_{\{\mu_2}\cdots p_{\mu_j}\epsilon_{\mu_1\}\lambda\xi\zeta}p_\xi P_\zeta p_\tau+p_{\{\mu_2}\cdots p_{\mu_j}\epsilon_{\mu_1\}\tau\xi\zeta}p_\xi P_\zeta p_\lambda,
\end{split}
\end{equation*}
\begin{equation*}
\begin{split}
\mathcal{T}_{\mu_1\cdots\mu_j\lambda\tau}^{9}=&-(2P\cdot p+\eta_1P^2-\eta_2P^2)p_{\{\mu_2}\cdots p_{\mu_j}\epsilon_{\mu_1\}\lambda\tau\xi}p_\xi\\
&+P\cdot(\eta_2p-\eta_1p+2\eta_1\eta_2P)p_{\{\mu_2}\cdots p_{\mu_j}\epsilon_{\mu_1\}\lambda\tau\xi}P_\xi\\
&+p_{\{\mu_2}\cdots p_{\mu_j}\epsilon_{\mu_1\}\lambda\xi\zeta}p_\xi P_\zeta P_\tau+p_{\{\mu_2}\cdots p_{\mu_j}\epsilon_{\mu_1\}\tau\xi\zeta}p_\xi P_\zeta P_\lambda,
\end{split}
\end{equation*}
\begin{equation*}
\begin{split}
\mathcal{T}_{\mu_1\cdots\mu_j\lambda\tau}^{10}=&-P^2p_{\{\mu_2}\cdots p_{\mu_j}\epsilon_{\mu_1\}\lambda\tau\xi}p_\xi+(P\cdot p)p_{\{\mu_2}\cdots p_{\mu_j}\epsilon_{\mu_1\}\lambda\tau\xi}P_\xi\\
&-p_{\{\mu_2}\cdots p_{\mu_j}\epsilon_{\mu_1\}\lambda\xi\zeta}p_\xi P_\zeta P_\tau+p_{\{\mu_2}\cdots p_{\mu_j}\epsilon_{\mu_1\}\tau\xi\zeta}p_\xi P_\zeta P_\lambda,
\end{split}
\end{equation*}
\begin{equation*}
\begin{split}
\mathcal{T}_{\mu_1\cdots\mu_j\lambda\tau}^{11}=&(p^2+\eta_1P\cdot p-\eta_2P\cdot p-\eta_1\eta_2P^2)p_{\{\mu_3}\cdots p_{\mu_j}g_{\mu_1\lambda}\epsilon_{\mu_2\}\tau\xi\zeta}p_\xi P_\zeta\\
&-p_{\{\mu_2}\cdots p_{\mu_j}\epsilon_{\mu_1\}\tau\xi\zeta}p_\xi P_\zeta(p-\eta_2P)_\lambda,
\end{split}
\end{equation*}
\begin{equation*}
\begin{split}
\mathcal{T}_{\mu_1\cdots\mu_j\lambda\tau}^{12}=&(p^2+\eta_1P\cdot p-\eta_2P\cdot p-\eta_1\eta_2P^2)p_{\{\mu_3}\cdots p_{\mu_j}g_{\mu_1\tau}\epsilon_{\mu_2\}\lambda\xi\zeta}p_\xi P_\zeta\\
&-p_{\{\mu_2}\cdots p_{\mu_j}\epsilon_{\mu_1\}\lambda\xi\zeta}p_\xi P_\zeta(p+\eta_1P)_\tau,
\end{split}
\end{equation*}
where $\{\mu_1,\cdots,\mu_j\}$ represents symmetrization of the indices $\mu_1,\cdots,\mu_j$.

\bibliographystyle{apsrev}
\bibliography{ref}

\end{document}